\documentclass[runningheads]{llncs}
\usepackage[T1]{fontenc}
\usepackage{graphicx}
\usepackage{booktabs}
\usepackage{xcolor}
\usepackage{caption}
\usepackage{subcaption}
\usepackage{listings}
\usepackage{bbding}

\newcommand{\lmul}{\texttt{LMUL}}

\begin{document}
\title{Closer in the Gap: Towards Portable Performance on RISC-V Vector Processors}

\titlerunning{Performance on RISC-V RVV}

\author{Ruimin Shi\inst{1} \orcidID{0009-0003-4387-367X}\and
Maya Gokhale\inst{2} \orcidID{0000-0003-4229-5735}\and
Pei-Hung Lin\inst{2} \orcidID{0000-0003-4977-814X}\and
Xavier Teruel\inst{3} \orcidID{0000-0001-5181-7545}\and
Ivy Peng\inst{1} \orcidID{0000-0003-4158-3583} \Envelope
}
\authorrunning{Shi, R., et al. }
%
\institute{KTH Royal Institute of Technology, Sweden \email{\{ruimins, ivybopeng\}@kth.se} \and
Lawrence Livermore National Laboratory, USA \email{\{gokhale2, lin32\}@llnl.gov} \and
Barcelona Supercomputing Center, Spain \email{xavier.teruel@bsc.es}}

\maketitle

\begin{abstract}
The RISC-V Vector Extension~(RVV) is a cornerstone for sustaining compute throughout in scientific and machine learning workloads. However, compiler support and performance monitoring on RVV~1.0 hardware remain actively evolving. In this work, we design a suite of assembly microbenchmarks to calibrate performance counters on the BananaPi-F3 RVV hardware. Using the benchmarks, we find that predication overhead and stride load pose performance bottlenecks that current compiler cost models do not yet fully capture. Evaluating GCC 15 and Clang 21 across six scientific and ML applications, GCC 15 produces more stable and efficient code generation, outperforming Clang 21 in four of six benchmarks. Clang 21 achieves superior performance only in SGEMM and DGEMM, attributable to more aggressive instruction reduction as confirmed by our validated perf counters. Default LMUL selection proves near-optimal in most cases, though GCC 15 shows greater potential for performance gains through larger LMUL tuning than Clang 21. Finally, we contribute an RVV backend for Google's Qsim quantum circuit simulator implemented using RVV intrinsics, where GCC 15 again outperforms LLVM 21 on this complex, real-world workload.

\end{abstract}

\section{Introduction}
The RISC-V Instruction Set Architecture (ISA) has attracted increasing interest from both industry and academia for its open, modular design and suitability for workload-specific customization~\cite{ramirez2020risc,brown2023risc,lee2023test,perotti2022new}. A cornerstone of RISC-V's relevance to High-Performance Computing is the \emph{RISC-V Vector Extension}~(RVV). Many HPC workloads, including ML training and inference, scientific simulations, and signal processing, are built around data-parallel algorithms that benefit directly from wide vector execution. Unlike traditional SIMD with fixed vector lengths in ISA, RVV adopts a \emph{Vector-Length-Agnostic}~(VLA) model where the active vector length can be determined at runtime. Another popular VLA realization is ARM's scalable vector extension (SVE), which was first implemented in Fujitsu's A64FX processor powering the Fugaku supercomputer, and widely adopted in recent NVIDIA Grace and AWS Graviton processors~\cite{shi2025arm}. 

For achieving portable performance on various RVV hardware, compiler support and performance profiling are both important. Auto-vectorization compiler support is the key enabler for portable performance on broad RVV hardware. At the fast pace of new RVV hardware emerging, compiler support allows applications to harness different vector hardware without rewriting applications. However, compared to vector length-specific SIMD, RVV's VLA model imposes new challenges on the compiler support for auto-vectorization. One particular challenge is that under uncertainty about the hardware vector width at compilation time, a compiler must generate correct \texttt{vsetvl} sequences that set active vector length at runtime, select an appropriate value for the vector length multiplier (LMUL), and handle tail elements and masks. Addressing these difficulties, GNU and LLVM are both under active development to ramp up RVV support.

Prior works have characterized compiler support for RVV, including comparative evaluations of GCC~14 and LLVM~19 on ratified RVV~1.0 hardware~\cite{adit2022performance,lee2023test,carpentieri2025performance}. However, both compilers have since undergone substantial RVV-specific development. For instance, GCC~15 introduced early-break loop vectorization, corrected memory misaligned access, and dynamic LMUL selection. LLVM~21 added multiple hardware-specific scheduling models for recent RVV hardware. An up-to-date evaluation of these compiler advances is therefore timely and necessary. Meanwhile, performance monitoring on RISC-V hardware remains significantly less mature than on x86 or AArch64~\cite{banchelli2025risc}. The RISC-V specification mandates only two HPM counters, \texttt{cycle} and \texttt{instruction}, leaving other events to custom-specific implementation. Therefore, validating the availability and reliability of performance monitoring counters remains necessary for optimizing performance on RVV hardware. In this paper, we make the following contributions:
\begin{itemize}
  \item We design a suite of assembly microbenchmarks targeting key RVV~1.0 arithmetic and data access operations, and validate RVV-specific hardware performance events in \textit{perf} on RVV hardware.
  \item We evaluate the enhanced support for RVV autovectorization in GCC15 and LLVM21 in six HPC and ML proxy applications on two RVV1.0 hardware.
  \item We quantify that predication masks introduce 35\% overhead w.r.t unmasked \texttt{vsetvl} for tailing elements, and strided vector ld/st instructions incur up to 4$\times$ the cost of unit-stride equivalents
  \item We contribute an RVV backend for a production quantum circuit simulator Qsim using RVV intrinsics and find that GCC 15 outperforms LLVM 21 on this complex, real-world workload. 
\end{itemize}

\section{Background}
\textbf{RVV Processors and VLA.} The RVV version v1.0~\cite{rvvspec} was ratified in 2021 and since then, hardware implementation has gained traction in vendors and undergone active development~\cite{perotti2022new}. RVV supports the VLA programming model, similar to ARM scalable vector extension (SVE). Different from vector length specific (VLS), such as x86 AVX and ARM NEON, which require the vector length known at compilation time, VLA allows the same binary to be portable across RVV hardware with different vector length (VLEN). RVV defines 32 VLEN-bit long vector registers. Recent implementations support VLEN from 128 up to 16384~\cite{perotti2022new}. Through dynamic register typing, RVV tag data element types and size to vector registers instead of instructions, to enable a fixed ISA supporting growing VLEN. The active vector length (VL) is determined at runtime via the \texttt{vsetvl} instruction and depends on the memory allocated to vector registers in hardware implementation, the number of activated vector registers, and the selected element width (SEW). LMUL is a multiplier of values 1/8, 1/4, 1/2, 1, 2, 4, or 8, which groups multiple vector registers into one `logical' vector so that when LMUL is larger than 1, more elements can be executed in one instruction. Combining these architectural parameters, the theoretical maximum number of elements that can be operated on with a single vector instruction is $VL = \frac{LMUL*VLEN}{SEW}$. RVV also provides the masking mechanism to support predicated vector execution and non-unit stride and indexed gather/scatter operations for irregular loop patterns.


\textbf{Compiler Support for VLA model.} RVV intrinsics and autovectorization are the primary mechanisms to translate RVV's architectural portability into practical performance portability in applications. Both GCC and LLVM have offered code generation compliant with RVV 1.0 specification. A capable compiler can generate code that automatically adapts to the underlying vector hardware at runtime, removing the need for platform-specific hand tuning required by VLS. Both GCC and LLVM have invested heavily in RVV~1.0 support.

The RVV~1.0 C intrinsics provide developers with direct access to individual vector instructions from the C/C++ source code, including control over SEW, LMUL, VL and masks. Intrinsics thus represent the performance ceiling against which compiler-generated autovectorized outcomes can be measured. For irregular access patterns, hand-written intrinsic code consistently outperforms compiler-generated code in earlier LLVM Clang releases.~\cite{adit2022performance}.

Compiler autovectorization transforms scalar loops into vector code automatically, selecting appropriate LMUL values, emitting \texttt{vsetvl} instructions~\cite{carpentieri2025performance}. 
GCC~14 and Clang~19 were the first GNU and LLVM releases with production-quality RVV autovectorization and full C intrinsics. GCC~15 extends this with saturating arithmetic(\texttt{vssubu.vv}, \texttt{vnclipu.wi}), early-break vectorization (loops with conditional \texttt{break} statements), and an improved \texttt{-O2} cost model. 
LLVM's RISC-V backend has received intensive investment since then, in areas including VL tail folding and split register allocation between the RVV and scalar register files. LLVM~21 advances by extending the RISC-V backend to support a rich set of CPU models, including the SpacemiT scheduling model. 

\section{Methodology}
We develop a set of hand-written RVV~1.0 assembly microbenchmarks \footnote{https://github.com/KTH-ScaLab/rvv-evaluation} for assessing key RVV~1.0 instructions and for verifying \textit{perf} performance profiling counters. Leverage those validated \textit{perf} counters, we compare and analyze the performance difference of GCC~15 and Clang~21 in autovectorizing six representative proxy applications from HPC and ML domains on two RVV~1.0 testbeds. 

\textbf{Assembly Microbenchmarks.} Since compiler-generated code may conflate micro-architectural behaviors with compiler decisions, we use a set of assembly microbenchmarks that issue precisely controlled sequences of RVV~1.0 instructions with explicit control of operand types and register configurations. We leverage the assembly codes to establish the performance ceiling in terms of raw throughput on the target testbed. Each assembly benchmark consists of a hand-written inner loop repeating a target RVV instruction for $10^8$ times. The operands are pre-staged in vector registers to eliminate cold-start effects. Dependency among successive instructions is broken to expose peak issue throughput. If VL$<$VLEN or mask operations are used, \texttt{vta} and \texttt{vma}, configured by \texttt{vsetvli}, will control the behavior of destination tail elements and inactive masked-off elements, respectively. 
We opt for the agnostic policy when preserving the tail/masked-off elements in the destination vector register is not required.

The first set of benchmarks targets memory instructions to support common access patterns. Unit-stride loads and stores (\texttt{vle}\emph{N}\texttt{.v} / \texttt{vse}\emph{N}\texttt{.v} \texttt{vd, (rs1)}) are used for sequential access patterns. Non-uniform access patterns are realized through two options: (1) strided loads and store (\texttt{vlse}\emph{N}\texttt{.v} / \texttt{vsse}\emph{N}\texttt{.v} \texttt{vd, (rs1)}), and (2) through masked vector loads (\texttt{vle}\emph{N}\texttt{.v} / \texttt{vse}\emph{N}\texttt{.v} \texttt{vd, (rs1), v0.t}) with an alternating mask pattern \texttt{1010...} stored in \texttt{v0.t}. Indexed gather and scatter operations (\texttt{vluxei}/ \texttt{vsuxei}) are not evaluated because they have the same access behavior as \texttt{vlse}/\texttt{vsse} in the fixed-stride case, but introduce the additional index register and are seldom selected by the compiler auto-vectorization. The second set of benchmarks targets floating-point and integer arithmetic instructions. Assembly codes target vector addition (\texttt{vfadd.vv}), multiplication (\texttt{vfmul.vv}), fused multiply-add (\texttt{vfmacc.vv}), and division (\texttt{vfdiv.vv}) in FP16 (\texttt{e16}), FP32 (\texttt{e32}), and FP64 (\texttt{e64}), and their integer counterpart instructions at 8-bit, 16-bit, 32-bit, and 64-bit element widths.   


\textbf{Performance Counter Calibration with Assembly Benchmarks.} Unlike mature architectures such as x86 or AArch64, where performance counters are thoroughly documented, the hardware performance counters on RVV hardware through the Linux \texttt{perf} subsystem are implementation-defined~\cite{banchelli2025risc}. Therefore, before using counter-derived metrics for analyzing RVV vectorization efficiency, we calibrate which events are reliable on this specific microarchitecture. In particular, we collect seven \texttt{perf} events: total retired instruction, vector instruction, vector ld/st instruction, FP ld/st instruction, and FP instruction. Table~\ref{tab:inst_stats} summarizes the measurement of these events in 10 assembly benchmarks, where the target instruction sequences are known from assembly codes and their counts are presented in \textit{Ref ins}. Take the \texttt{vfadd.vv} benchmark for example, a loop issuing N \texttt{vfadd.vv} must retire N vector FP instructions and a small number of extra scalar loop-control instructions. Thus, the total retired instructions as reported in \textit{Retired ins.} should be close to but slightly higher than \textit{Ref ins}. 

A counter is deemed reliable if its reported value approximates the reference within a tolerance of 5\% across five independent runs. This threshold is chosen to accommodate minor perturbations from interrupt handling and context-switch overhead, a known limitation in RVV performance monitoring ~\cite{banchelli2025risc}. The retired instruction achieves 98.4-99\% accuracy on all 10 benchmarks. Similarly, vector ld/st and FP ld/st instructions only show $<10^{-5}\%$ error. Counters whose reported values are beyond the tolerance are unreliable and thus excluded in the profiling analysis in Sections~\ref{sec:proxy}. Vector instruction, which records all vector instructions, is expected to be close to 0 in \texttt{fadd} and \texttt{fmadd} benchmark, shows an error of approximately $50\%$. A similar issue appears in FP instructions \texttt{vfadd.vv} and \texttt{vmacc.vv}, where the expected value is also near zero, but the real value shows an error of $100\%$.


\begin{table}[t]
\centering
\caption{Perf event statistics for different assembly instructions.}
\label{tab:inst_stats}
\setlength{\tabcolsep}{4pt}
\resizebox{\linewidth}{!}{
\begin{tabular}{lllllllll}
\toprule
Assembly bench. & Ref \#ins. & Retired ins. & Vec. ld ins.& Vec. st ins.& Vec. ins. & FP ld ins.& FP st ins. & FP ins. \\
\midrule
flw      & $1.28\times10^{10}$& $1.31\times10^{10}$& 16 & 16 & 55 & $\mathbf{1.28\times10^{10}}$ & 696 & $1.28\times10^{10}$ \\
lw       & $1.28\times10^{10}$ & $1.31\times10^{10}$ & 16 & 16 & 53 & 736 & 855 & 1613 \\
vle.vv   & $1.28\times10^{10}$ & $1.31\times10^{10}$ & $\mathbf{1.28\times10^{10}}$ & 17 & $\mathbf{1.28\times10^{10}}$ & 662 & 738 & 2443 \\
fsw      & $1.28\times10^{10}$ & $1.31\times10^{10}$ & 16 & 16 & 55 & 601 & $\mathbf{1.28\times10^{10}}$ & $1.28\times10^{10}$ \\
sw       & $1.28\times10^{10}$ & $1.31\times10^{10}$ & 16 & 16 & 53 & 736 & 855 & 1613 \\
vse.vv   & $1.28\times10^{10}$ & $1.32\times10^{10}$ & 1108 & $\mathbf{1.28\times10^{10}}$ & $\mathbf{1.28\times10^{10}}$ & 662 & 738 & 2443 \\ \hline
vfadd.vv & $1.28\times10^{10}$ & $1.30\times10^{10}$ & 124 & 17 & $\mathbf{1.28\times10^{10}}$ & 652 & 726 & \textcolor{black}{\textbf{$2.56\times10^{10}$}} \\
vmacc.vv & $1.28\times10^{10}$ & $1.30\times10^{10}$ & 144 & 17 & $\mathbf{1.28\times10^{10}}$ & 632 & 723 & \textcolor{black}{\textbf{$2.56\times10^{10}$}} \\
fadd     & $1.28\times10^{10}$ & $1.30\times10^{10}$ & 16 & 16 & \textcolor{black}{\textbf{$6.40\times10^{9}$}} & 770 & 856 & $1.28\times10^{10}$ \\
fmadd    & $1.28\times10^{10}$ & $1.30\times10^{10}$ & 16 & 16 & \textcolor{black}{\textbf{$6.40\times10^{9}$}} & 770 & 856 & $1.28\times10^{10}$ \\
\bottomrule
\end{tabular}}
\vspace{-1.5em}
\end{table}


\textbf{Application Benchmarks.} To evaluate auto-vectorization performance under realistic workload conditions, we select six proxy applications representative of dominant computational and memory patterns in scientific and ML workloads. Stream measures sustained memory bandwidth through simple element-wise kernels. SpMV exhibits irregular memory access via sparse matrix-vector multiplication, where indirect indexing and poor spatial locality make it latency-bound. DGEMM and SGEMM implement double- and single-precision dense matrix multiplication, respectively, whose loop structure is regular and compute-bound, representing ideal cases for autovectorization. YOLOv3 and AlexNet represent CNN inference workloads that mix regular convolution with non-linear activations and pooling. Together, these applications cover diverse patterns and exercise the various data precision characterized in the assembly microbenchmarks. Further, we evaluate a production-level application for simulating quantum computers developed by Google, called QSIM, to investigate the RVV compiler support for both auto-vectorization and RVV intrinsics. For this, we also provide a manually ported QSIM using RVV intrinsics for comparison \footnote{https://github.com/KTH-ScaLab/qsim\_rvv}.

\textbf{Compiler and Testbed Setup.} GCC15.1.0 and Clang21.1.1 are used. All compiler configurations are augmented with "-O3/-Ofast -fopenmp".
The first version of compilation, \textit{non-vec}, uses flag "-march=rv64gc -fno-tree-vectorize" in GCC15 and "-mllvm -scalable-vectorization=off" in Clang21, to disable the vectorization. The second version, \textit{auto-vec}, uses flag "-march=rv64gcv\_zfh\_zvfh" in GCC15 and "-march=rv64gcv\_zfh\_zvfh -mllvm -scalable-vectorization=on" in Clang21. This version enables the autovectorization and uses the default \lmul{} setting in each compiler. In a variant, we also specify \lmul{}, using "-mrvv-max-lmul" in GCC15 and "-mllvm -riscv-v-register-bit-width-lmul" in Clang21. 

The testbeds consist of Milk-V Jupiter and BananaPi BPI-F3, both are based on the Spacemit(R)~X60 CPU with 8 cores implementing $2\times 256$-bit RVV1.0 support, running at 1.6~GHz (BananaPi) and 1.8~GHz (Jupiter) frequency. The memory hierarchy includes 32~KB L1 instruction and 32~KB L1 data caches per core, and a shared $2\times 512$~KB L2 caches. The systems are equipped with 16~GB of LPDDR4X (Milk-V Jupiter) and LPDDR4 (BPI-F3) main memory. We use the perf in Linux kernel 6.1.15 and kernel 6.6.63 on the BSC testbed.

\section{Understanding Basic RVV Instructions}
\begin{figure}[ht]
    \vspace{-1.5em}
    \centering
    \includegraphics[width=0.8\linewidth]{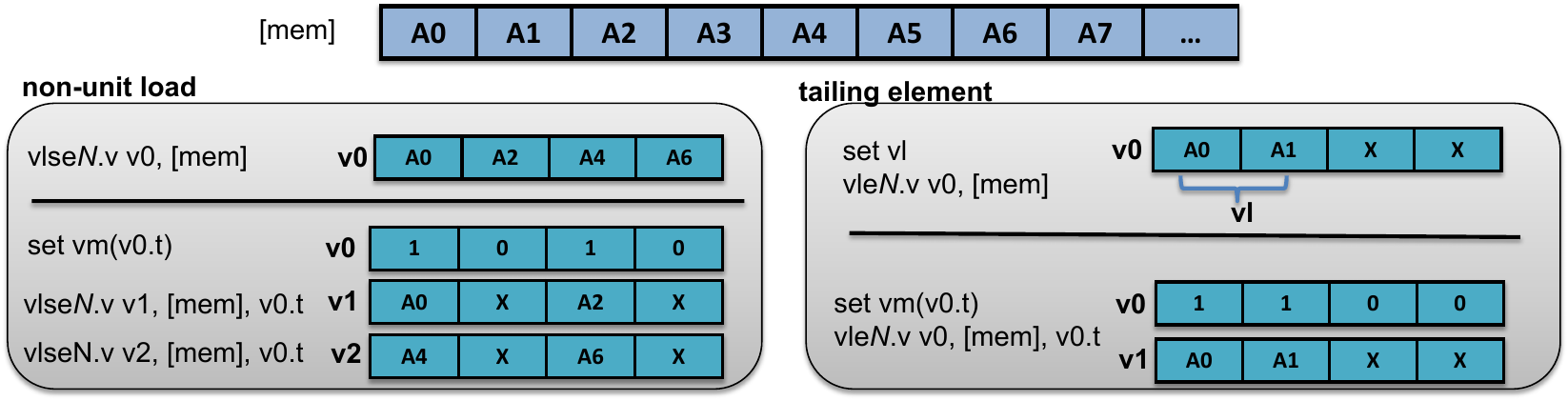}
    \caption{The diagram of non-unit and tailing vector load}
    \label{fig:diagram_vlse_vle}
    \vspace{-1.5em}
\end{figure}

We start our evaluation with two categories of vector instructions, i.e., data access instructions and arithmetic instructions. For ld/st instructions, we investigate three access patterns, covering sequential, non-unit, and tailing elements. While the sequential ld/st can be achieved with the \texttt{vle/vse} instructions, non-unit and tailing elements can be achieved in different ways, as illustrated in Figure~\ref{fig:diagram_vlse_vle}. Thus, we design the assembly benchmark for each variant to identify the recommended choice for programmers.

For a sequential load pattern, unit-stride load \texttt{vle8} achieves the highest throughput of 28.4 Gops/s on Jupiter (25.2 Gops/s on BPI-F3), and its \texttt{16|32|64}-bit variants scale the throughput proportionally to 14.2,7.1,3.55 Gops/s, respectively. This throughput approaches the hardware's theoretical peak closely. 

For non-unit load patterns, we focus on the stride access patterns. For instance, for stride=2, \texttt{mem[0], mem[2], mem[4], $\cdots$, mem[2*i]} should be accessed, shown in  Figure \ref{fig:diagram_vlse_vle}. We design three benchmarks to compare different instructions to achieve the same access pattern. In the first benchmark, vector stride load instruction \texttt{vlse} is used to gather elements from memory with a stride of 2 into a vector register. The second benchmark uses united load \texttt{vle} to access contiguous elements from \texttt{mem} and a mask register (1010) to only enables the corresponding positions of required elements \texttt{0, 2, 4, $\cdots$} in the destination vector register. To load the same number of elements as a single \texttt{vlse}, this method requires two masked \texttt{vle} instructions. Moreover, the configuration instructions to generate masks are added at initialization stage. All additional overheads are included in the throughput measurements. Finally, the third benchmark adds a reference to a scalar implementation, where scalar loads are used to read elements with a constant stride from memory to scalar registers.

\begin{figure}[ht]
\vspace{-1em}
\centering
\begin{minipage}{0.45\linewidth}
    \centering
    \includegraphics[width=\linewidth]{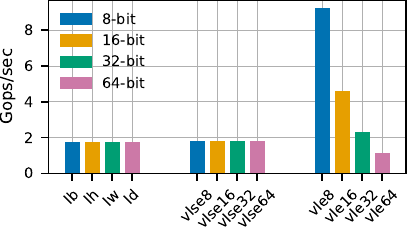}
    \caption{The peak throughput of non-uniform load instructions on Jupiter.}
    \label{fig:ld_ins}
\end{minipage}
\hfill
\begin{minipage}{0.5\linewidth}
    \centering
    \includegraphics[width=0.9\linewidth]{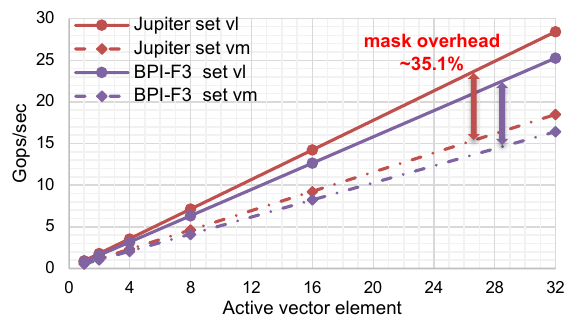}
    \caption{Compare the performance of tailing elements via setvl and mask operations on BPI-F3 and Jupiter.}
    \label{fig:mask_overhead}
\end{minipage}
\vspace{-1.5em}
\end{figure}
Figure \ref{fig:ld_ins} presents the results of the three benchmarks for stride loads on Jupiter. The results show \texttt{vlse} and scalar are consistent at 1.78 Gops/s in any variants. \texttt{vle} with mask achieves the best performance of 9.2 Gops/s at 8-bit precision and the throughput scales up with reduced data precision.
In 64-bit data precision, the throughput of the \texttt{vle} with mask is 0.65$\times$ lower than \texttt{vlse}.  
Store has identical performance as load at all instructions, except \texttt{vsse32} only achieving 1.2 Gops/s, thus omitted from Figure \ref{fig:ld_ins}. 


RVV provides a unique way of handling tail elements by setting its active vector length VL, in addition to the masked predicate operations. To guide the selection between these two options, we design two assembly benchmarks to compare these two ways of handling irregular vector lengths. Shown in Figure \ref{fig:diagram_vlse_vle}(b), we load \texttt{vl} elements with useless tail element in destination vector register, where \texttt{vl}< \texttt{VLMAX}.   The first benchmark uses the \texttt{vset\{i\}vl\{i\}} instruction to change the VL. The second benchmark achieves the same goal by setting the mask register \texttt{v0.t}. Figure~ \ref{fig:mask_overhead} presents the performance comparison of these two methods. Setting \texttt{vl} reaches about 28.4 Gops/s and 25.2 Gops/s at 32 active elements, whereas mask achieves only 18.5 Gops/s and 16.4 Gops/s on Jupiter and BPI-F3, respectively. A constant 35.1\% performance loss in masked operations at any active vector elements. The mask configuration introduces negligible cost, as only three additional instructions are required in initialization. The observed overhead mainly comes from vector execution, representing the inefficiency in executing the masked vector instructions in vector units.


\begin{figure}[bt]
    \centering
    \includegraphics[width=0.9\linewidth]{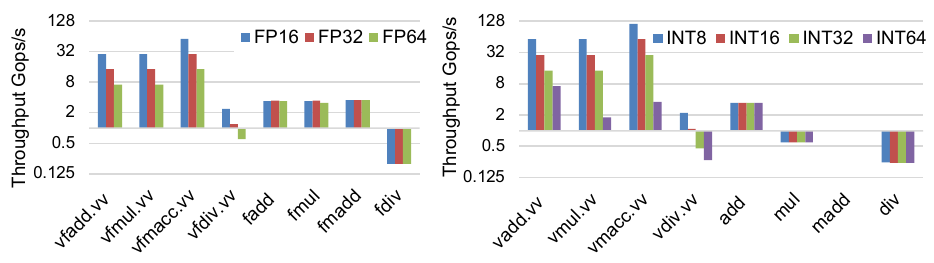}
    \caption{The peak throughput of selected vector and scalar arithmetic instructions in FP16/32/64 and INT8/16/32/64 on Jupiter.}
    \label{fig:arith_ins}
    \vspace{-1.2em}
\end{figure}

Evaluating peak arithmetic throughput is essential to establish the performance ceiling and determine whether the $2\times$256-bit vector pipelines deliver the theoretical throughput across different element sizes. In this evaluation, we focus on floating-point (FP16|32|64) and integer (INT8|16|32|64) operations important for scientific and machine learning workloads. We design each microbenchmark consisting of a tight, dependency-free loop of a single RVV~1.0 instruction, e.g., (\texttt{v(f)add.vv}, \texttt{v(f)mul.vv}, \texttt{v(f)macc.vv}, \texttt{v(f)div.vv}), issued at different LMUL. The benchmark preloads operands into vector registers to eliminate memory latency and repeats $4 \times 32$ instructions with \texttt{v0-v32} as destination per iteration and $10^8$ iterations to reach steady operation throughput. We also measure the scalar counterparts (\texttt{(f)add}, \texttt{(f)mul}, \texttt{(f)madd}) identically to provide the baseline for quantifying the vector speedup. Figure \ref{fig:arith_ins} summarizes the measured throughput. 

The result reveals that the fused multiply-add vector operation achieves the highest throughput of all measured instructions, with FP16 reaching 57.5~Gops/s, 28.7~Gops/s for FP32, and 14.4~Gops/s for FP64, consistent with the expected doubling of element count per vector register as precision is halved. The fused multiply-add \texttt{vfmacc.vv} at FP32 nearly matches the throughput of \texttt{vfadd.vv} and \texttt{vfmul.vv}, demonstrating that the hardware can issue fused operations at full vector throughput. This is favourable for BLAS-like kernels that are dominated by DAXPY and GEMM patterns. The scalar counterparts, \texttt{fadd}, \texttt{fmul}, and \texttt{fmadd} instructions achieve only $3.1$-$3.5$~Gops/s regardless of precision, indicating $16\times$ throughput advantage of the vector variants for FP16. For INT arithmetic instructions, \texttt{vmul.vv} scales proportionally 51.1, 25.6, 12.8 Gops/s with INT8-32 and only 1.6 Gops/s for INT64, while scalar \texttt{mul} only achieves a constant at 0.53 Gops/s, even worse than \texttt{fmul}, indicating the inefficient scalar multiplier support in ALU. Both \texttt{vdiv.vv} and scalar \texttt{div} show very low throughput(0.22- 1.97 Gops/s), reflecting the high latency of the divider. This encourages the programmers and compiler optimization to replace division with shift or multiplication if possible.

\section{GCC15 and Clang21 Support in Proxy Apps}
\label{sec:proxy}
While the assembly benchmark helps understand performance ceiling, evaluating proxy applications from scientific and ML workloads is necessary to assess whether autovectorization translates into application level speedups. We evaluate six proxy applications compiled with GCC~15 and Clang~21 into non-vectorized and autovectorized binaries. We report all speedups normalized against the non-vectorized GCC~15 baseline. We evaluate performance on both BPI-F3 and Jupiter and observe consistent conclusions across the two platforms. Therefore, unless otherwise stated, the following results presented are from Jupiter.


\begin{figure}[bt]
\centering
\begin{subfigure}{0.48\columnwidth}
    \centering
    \includegraphics[width=\linewidth]{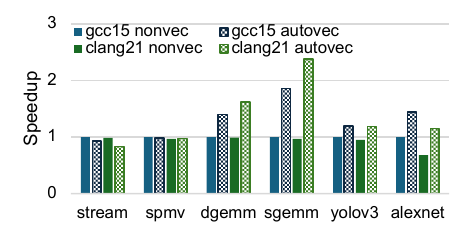}
    \caption{Runtime speedup}
    \label{fig:speedup_jupiter}
\end{subfigure}
\hfill
\begin{subfigure}{0.49\columnwidth}
    \centering
    \includegraphics[width=\linewidth]{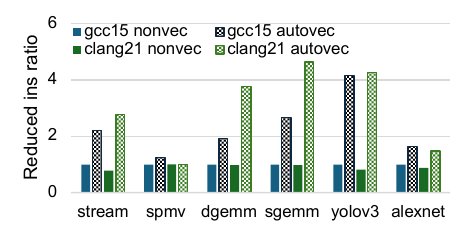}
    \caption{Reduced instruction ratio}
    \label{fig:reduce_ins_jupiter}
\end{subfigure}
\caption{The performance by GCC~15 and Clang~21 autovectorization across six proxy apps on Jupiter, normalized by the GCC~15 nonvec baseline}
\vspace{-1.6em}
\end{figure}

Figure~\ref{fig:speedup_jupiter} presents an overview of achieved speedup by GCC15 and Clang21 on Milk-V Jupiter. The autovectorization benefits are highly application-dependent. Clang~21 achieves more speedup than GCC~15 in sgemm and dgemm, where sgemm gains approximately $2.4\times$ speedup with Clang~21 and around $1.85\times$ with GCC~15. While both compilers can effectively vectorize gemm's regular inner loop, Clang~21 yields better speedup probably because of its SpacemiT X60 specific instruction scheduling model and LMUL selection. In the YOLOv3 and AlexNet benchmarks, GCC~15 achieves better speedup than Clang~21. For instance, AlexNet shows approximately $1.45\times$ speedup with GCC~15 but only $1.2\times$ speedup with Clang~21. Stream and SpMV, two memory-bound benchmarks, show no autovectorization benefit from both compilers. While GCC~15 achieves the same performance as the non-vectorized version, Clang~21 even decreases the performance to be worse than the non-vectorized version in stream, i.e., the Clang~21 vectorized stream is about $20\%$ slower than Clang~21 non-vectorized version. A notable observation from comparing the non-vectorized baseline is that Clang~21 is consistently slightly below 1.0, suggesting that GCC~15 generates better scalar codes. Figure~\ref{fig:reduce_ins_jupiter} reveals that instruction reduction ratio is a strong predictor of speedup, but the relationship is application-dependent. SGEMM and YOLOv3 show the largest reductions in instructions, with Clang~21 autovec, retiring $4.7\times$ and $4.3\times$ fewer instructions, respectively than the GCC~15 scalar baseline. While the reduction in instructions in SGEMM and YOLOv3 confirms that wide vector operations successfully amortize computation across multiple elements, stream also exhibits a large instruction reduction ($2.2\times$ for GCC~15 and $2.8\times$ for Clang~21) but without a corresponding speedup, which is as expected for its memory-bandwidth nature. The DGEMM and AlexNet reductions are moderate, i.e., approximately $2.0\times$ and $1.6\times$ for GCC~15 autovec. They correlate with their moderate speedups.

SpMV shows nearly no instruction reduction under either compiler in RVV. For comparison, we evaluate the same kernel on another VLA architecture, ARM SVE, where a 1.99× instruction reduction by GCC auto-vectorization is observed on a 128-bit SVE Neoverse V2 core. This discrepancy suggests that the automatic vectorization support for RVV remains less mature and requires further improvements in the compiler toolchains.

Comparing the two compilers, Clang~21 consistently reduces more instructions than GCC~15 across all vectorized applications, but this advantage is most effectively converted into speedup when the application is compute-bound. For memory-bound workloads, the instruction savings go largely unrealized as the LPDDR4X channel memory subsystem becomes the bottleneck. 

\begin{figure}[bt]
    \centering
    \includegraphics[width=\linewidth]{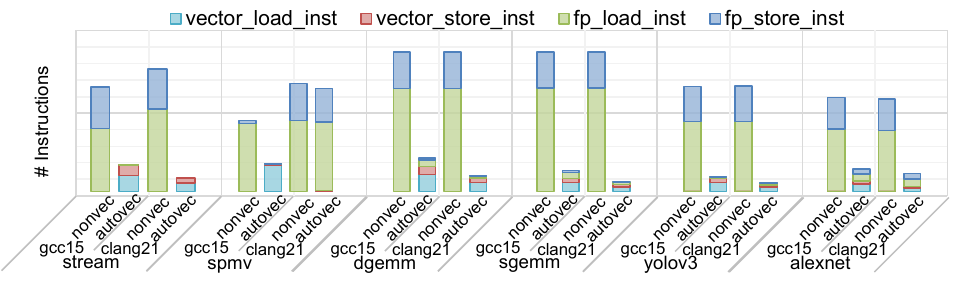}
    \caption{The breakdown load/store instructions in BPI-F3}
    \label{fig:Profiling}
    \vspace{-1.8em}
\end{figure}
\textbf{Profiling Analysis.} In this section, we leverage those reliable performance counters to decompose the retired instruction mix into vector load and store, floating-point load and store events to compare the instruction breakdown in different compiler-vectorized versions. The profiling result present in Figure~\ref{fig:Profiling} reveals that autovectorization dramatically reduces the total number of memory access instructions across all six applications, with the most reductions occurring in DGEMM and SGEMM. 

In SGEMM, the non-vectorized baseline is dominated by a large portion of FP ld/st instructions, which collapse to a small fraction under autovectorization, as each vector load can replace multiple scalar FP loads, indicating both GCC~15 and Clang~21 can effectively recognize such patterns. In Stream, the autovectorized version successfully eliminates most scalar load/store to a small residual of vector load and store instructions. However, although Clang~21 achieves a larger instruction reduction for STREAM, the observed performance is worse, likely caused by unsuitable memory access order after vectorization. As an in-order CPU model, the performance of Spacemit(R) X60 is sensitive to the static instruction scheduling by compilers. SpMV is the notable exception: both non-vectorized and autovectorized versions retain a similar instruction mix with a dominant FP load component and negligible vector memory instructions in Clang~21. Although vector load dominates in GCC~15 and obviously reduces the FP ld/st instructions, the total instructions are almost not reduced according to Figure \ref{fig:reduce_ins_jupiter}.
Further analyzing the compiler vectorized report, for the indexing memory access \texttt{vector[colIndex[col]]}, Clang~21 can not identify array bounds in SPMV, avoiding any parallelization optimization techniques. GCC~15 can vectorize part of the code using variable-length unit vector load, but additional missed optimizations associated with statement clobbers memory, avoiding further vectorization in fused-multiply-add.

Comparing GCC~15 and Clang~21, Clang~21 without vectorization generates more scalar memory instructions, likely due to differences in scalar loop unrolling and scheduling policies. Under autovectorization, however, Clang~21 reduces its instruction stack more aggressively than GCC~15 in DGEMM, SGEMM, and the ML workloads. The autovectorized Clang~21 bar for SGEMM approaches the smallest total instruction count of any configuration, consistent with its larger instruction-reduction ratio and higher speedup. For YOLOv3 and AlexNet, GCC~15 autovec and Clang~21 autovec produce nearly identical instruction mixes, which explains why their speedups converge for these workloads despite the larger scalar-code difference between the two compilers. 

\begin{figure}[bt]
\centering
\begin{subfigure}{0.48\columnwidth}
    \centering
    \includegraphics[width=\linewidth]{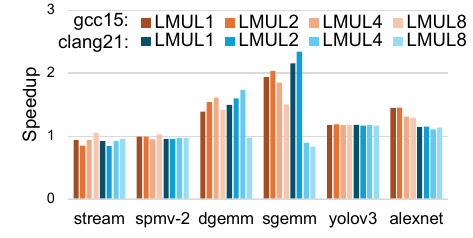}
    \caption{runtime speedup}
    \label{fig:speedup_LMUL_jupiter}
\end{subfigure}
\hfill
\begin{subfigure}{0.48\columnwidth}
    \centering
    \includegraphics[width=\linewidth]{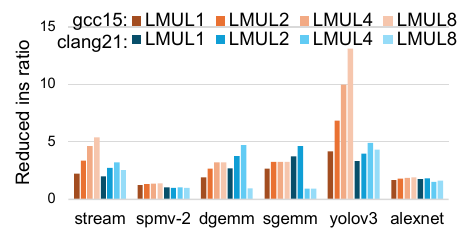}
    \caption{Reduced instruction ratio}
    \label{fig:reduce_ins_LMUL_jupiter}
\end{subfigure}
\caption{The impact of LMULs selection on Jupiter, normalized by GCC~15 nonvec}
\label{fig:LMUL}
\vspace{-1.6em}
\end{figure}

\textbf{Impact of LMUL in GCC15 and Clang21.} Although autovectorization defaults to a compiler-chosen LMUL value, the choice of LMUL is known to affect performance profoundly by trading register count against element throughput per instruction~\cite{lee2023test,carpentieri2025performance}. Higher LMUL values process more elements per cycle but reduce the number of available architectural vector registers from 32 (at LMUL=1) down to just 4 (at LMUL=8), trading off between instruction-level parallelism and register pressure. Therefore, we sweep LMUL in these proxy applications to determine whether the compiler's default LMUL selection is optimal for each workload and to guide users toward performance-portable choices when targeting the SpacemiT microarchitecture.

Figure~\ref{fig:LMUL} shows that the optimal LMUL setting is application- and compiler-dependent for performance. For SGEMM, Clang~21 at LMUL=2 achieves the highest speedup, approximately $2.35\times$. However, the performance degrades sharply at LMUL=8, to around $0.85\times$, below scalar performance, due to the high risk of register spilling. GCC~15 on SGEMM and DGEMM benefits from increasing LMUL up to $LMUL=4$, reaching approximately $2.0\times$ and $1.6\times$ respectively. One hypothesis is that the conservative unrolled and vectorized loop strategy in GCC~15 allows it to better tolerate the higher register pressure caused by larger LMUL. Stream and SpMV remain near or below $1.0\times$ across all LMUL values for both compilers. This is expected because their bottlenecks at memory bandwidth and irregular memory access latency, respectively, could be even exacerbated by widening vector register groups. For most applications, selecting LMUL=1 or LMUL=2 provides better performance. Larger LMUL values increase register pressure and may trigger register spilling, which significantly degrades performance. This observation suggests that the compiler’s default LMUL selection strategy is close to the optimal, and tuning the larger LMUL in GCC~15 can get more performance gains than Clang~21.


\begin{figure}[bt]
\centering
\begin{minipage}{0.48\columnwidth}
    \centering
    \includegraphics[width=\linewidth]{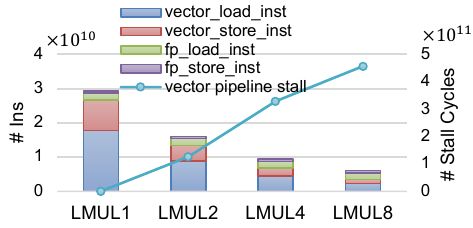}
    \caption{Yolov3 profiling analysis on the impact of LMULs}
    \label{fig:yolov3_LMUL_merged}
\end{minipage}
\hfill
\begin{minipage}{0.48\columnwidth}
    \centering
    \includegraphics[width=\linewidth]{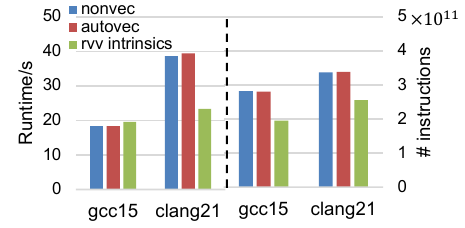}
    \caption{The comparison of Qsim across 3 versions using 8 cores}
    \label{fig:qsim_merged}
\end{minipage}
\vspace{-1.6em}
\end{figure}

Instruction reduction does not always translate into performance speedup, for example, YOLOv3 by GCC shows a proportional instruction reduction ratio with LMUL(13.1× at LMUL=8), closer to the theoretical value. However, the speedup of YOLOv3 and AlexNet is insensitive to LMUL for both compilers, remaining $1.2\times$, implying that their performance is constrained by some other factors other than VL. To understand the execution bottleneck of YOLOv3 in larger LMUL, we further break down its profiling results. We profile the instruction composition to verify the effectiveness of vectorization, shown in Figure \ref{fig:yolov3_LMUL_merged}. The profiling results show that scalar load/store instructions are largely replaced by vector load/store instructions, indicating that the kernel has been effectively vectorized. However, despite the reduction in total instruction and ld/st instructions, no speedup benefit is observed. To investigate the bottleneck, we profile the vector pipeline behavior. Figure \ref{fig:yolov3_LMUL_merged} reflects that the vector store pipeline cycles increase linearly with larger LMUL, rising from $2.3\times 10^8$ at LMUL1 to $4.6\times 10^{11}$ at LMUL8. This indicates that aggressive vector register grouping increases pressure on the vector pipeline, eventually exceeding its peak throughput. Consequently, instructions must wait for available vector pipeline resources to execute and commit, resulting in pipeline stalls.
\section{Qsim Quantum Circuit Simulator}

Google’s Qsim \cite{quantumaiteam} is a Schrodinger full state-vector simulator, which is highly optimized for FP32 arithmetic. Previous porting on ARM SVE backends achieves high performance~\cite{shi2026high}. 
Qsim uses interleaving layout to store the real and imaginary parts of the complex value for each state vector. The interleaved memory access pattern in Qsim makes compiler auto-vectorization difficult. To enable effective vectorization on RISC-V platforms, we port the implementation using RVV intrinsics via VLEN-adaptive memory layout adjustment and fine-grained vectorized loop by setting proper masks in operation to disable useless elements per vector instructions.

The autovec version shows nearly identical runtime to the nonvec version in GCC~15(18.3~s) and Clang~21(39.4~s) in Figure \ref{fig:qsim_merged} and no instruction reduction is observed, indicating that the compiler fails to effectively vectorize Qsim. The RVV intrinsics implementation significantly reduces the instruction count on both compilers, achieving around 1.4 $\times$ reduction ratio with GCC~15 and 1.3 $\times$  with Clang~21. However, the runtime with GCC~15 is even slightly longer than nonvec version, whereas Clang~21 achieves 1.6 $\times$ speedup compared to its own nonvec version. Despite the improvement from RVV intrinsics in Clang~21, its absolute runtime remains about 1.2 $\times$ slower than GCC~15, and the generated instructions is 1.2-1.3 $\times$ higher. Qsim results highlight the limitations of current compiler auto-vectorization for applications with complex memory access patterns as Qsim. GCC~15 performs better than Clang~21 in this real-world workload. Manual RVV intrinsics can effectively reduce instruction counts, but the achieved performance is strongly dependent on compiler code generation quality, requiring further optimization and compiler support for VLA architectures.

\section{Related Work}
Understanding and profiling the performance of RVV processors is critical for both architectural design and software optimization. Previous studies evaluated and analyzed the performance of a set of RISC-V processors equipped with RVV, ranging from high-performance computing~\cite{brown2023risc,lee2023test} to machine learning~\cite{gupta2023accelerating,garcia2025inference} workloads. For the feasibility and maturity of compiler support for auto-vectorization, several works evaluated the performance comparison between the GNU and LLVM compiler toolchains for RISC-V RVV across TSVC scientific loops and real applications~\cite{carpentieri2025performance,lai2025risc}. Adit et al.~\cite{adit2022performance} emphasized a significant performance gap between auto-vectorized and hand-optimized code for the RVV extension by examining LLVM’s support for both RVV and VLS-style Intel AVX and ARM Neon. Peccia et al.~\cite{peccia2025tensor} auto-tuned the vector instructions for AI workloads by integrating the scheduler into the TVM compiler.

Several studies explored the performance on specific RVV hardware via rewriting assembly and intrinsics~\cite{rvvbench,lin2024rewriting}. Since high-end RVV hardware is still emerging, evaluating on the cycle-accurate simulators such as GEM5~\cite{gupta2023accelerating,ramirez2020risc} is used to profile the detailed vector activities and explore the RVV design space. Ref~\cite{banchelli2025risc,vizcaino2025designing} compare three instrumentation methods: inline assembly CSR reads, Linux perf, and PAPI and proposed a Linux 6.10 kernel patch to toggle between perf-managed and direct CSR-based counter access. 

\section{Conclusion}
In this work, we first developed assembly benchmarks to calibrate vector-related \textit{perf} events on BananaPi RVV hardware. Evaluating the latest GCC~15 and Clang~21 support for RVV~1.0 auto-vectorization and intrinsics across six scientific and ML applications, we find that GCC~15 produces more stable and efficient code generation than Clang 21. Default LMUL selection proves near-optimal in most cases, though GCC 15 shows greater potential for performance gains through larger LMUL tuning than Clang 21. We further developed an RVV backend for Google's Qsim quantum circuit simulator using RVV intrinsics, where GCC 15 again outperforms LLVM 21 on this complex, real-world workload. Nevertheless, compared to mature compiler support for ARM SVE, both compilers require continued development to fully realize the performance potential of RVV 1.0. 

%
%
\begin{credits}
\vspace{-0.5em}
\subsubsection{\ackname} This work was performed under the auspices of the U.S. Department of Energy by Lawrence Livermore National Laboratory under Contract  DE-AC52-07NA27344 under LDRD Project 24-ERD-047. This research is supported by the European Commission under the Horizon project HIGHER (101189612). This research is supported by the RISC-V ExCALIBUR project at EPCC.
The authors have no competing interests to declare that are relevant to the content of this article.
\end{credits}

\bibliographystyle{splncs04}
\bibliography{main}
\end{document}